# Toroidal dipole response in a multifold double-ring metamaterial

Zheng-Gao Dong,[1,2] Peigen Ni,[1] Jie Zhu,[1] Xiaobo Yin,[1] and X. Zhang[1,3,*]

[1]*Nanoscale Science and Engineering Center, University of California, 5130 Etcheverry Hall, Berkeley, California 94720-1740, USA*
[2]*Physics Department, Southeast University, Nanjing 211189, China*
[3]*Materials Sciences Division, Lawrence Berkeley National Laboratory, 1 Cyclotron Road, Berkeley, California 94720, USA*
[*]*xiang@berkeley.edu*

**Abstract:** The toroidal response is numerically investigated in a multifold double-ring metamaterials at the antibonding magnetic-dipole mode (i.e., antiparallel magnetic dipoles in one double-ring fold). This intriguing toroidal resonance in metamaterials is considered as a result of the magnetoelectric effect due to the broken balance of the electric near-field environment. We demonstrate that the toroidal dipole response in metamaterials can improve the quality factor of the resonance spectrum. In viewing of the design flexibility on the double-ring geometry, such toroidal metamaterials will offer advantages in application potentials of toroidal dipolar moment.

## 1. Introduction

Metamaterials can realize a lot of intriguing physical phenomena due to the flexibilities in designing the constitutive meta-atoms and meta-molecules by structured elements. Some of these phenomena are inexistent in naturally occurring materials, such as the left-handed electromagnetic behavior and the cloaking [1,2]. Moreover, some of them are analogous to those existed in natural material systems, such as the chirality [3], the electromagnetically induced transparency (EIT) [4], and the celestial mechanics [5]. More recently, a metamaterial analogue of toroidal dipole response, represented by a closed loop of head-to-tail magnetic dipole distribution, was realized experimentally by a metamaterial composed of metallic split-ring array [6,7]. As is well known, electric and magnetic dipoles are two fundamental responses of an object in interaction with incident electromagnetic waves, corresponding to a charge oscillation and a circulating electric current, respectively. Higher-order responses, namely, multipoles like quadrupole and octupole, are usually weak as compared with the electric or magnetic dipole response. However, in contrast to these common dipole and multipole responses, a toroidal moment is not involved in the standard multipole expansion. It is produced by poloidal currents along the toroid surface, and was referred to as an anapole moment [8,9]. The elusive electromagnetic response of toroidal moment has attracted great attentions in nuclear, molecular, and ferroelectric physics because of its different characteristic from the fundamental electric and magnetic dipoles [10-12]. As a result, potential applications of the toroidal response were theoretically reviewed, based on the phase transition as well as the magnetoelectric effect [13]. Other applications for the toroidal response in nanostructured artificial materials have been reported recently in literatures, such as sensibility due to the high quality factor [6], circular dichroism and polarization controllability due to the optical activity [8], and negative refraction and backward waves due to the toroidal nature [9]. Moreover, optical nonlinearity enhancement was experimentally confirmed, attributed to the toroidal susceptibility that breaks both space-inversion and time-reversal symmetry [14].

Thereby, it is of interest to realize the toroidal moment by metamolecules, as an analogue of naturally occurring elementary particles and magnetic materials. Unfortunately, the toroidal dipole moment can neither be obtained in a metallic torus [15], nor be excited magnetically in a torus-like metamaterial by straightforwardly rotating a magnetic-dipole structure [6]. A typical result of such a treatment is just a linear superposition of individual magnetic dipoles, leading to a nonvanishing net magnetic dipole moment. In a recent work of ours, the dark mode of magnetic dipole resonance inherent to the split ring was verified to be excited electrically due to the asymmetric near-field environment around the split ring [16]. From this point of view, the toroidal dipole response in metamaterials [6,7] can be explained intuitively by a magnetoelectric effect (bianisotropy) attributed to the nonuniform near-field environment for each of the four wire loops. That is, the four loop edges parallel and near to the symmetry axis of the toroidal loop array (inner edges) are electrically unbalanced, in terms of the near field distribution, regarding to the other four loop edges parallel but far away from this axis (outer edges). Consequently, there are in-phase circulating currents on the loop surfaces flowing from inner edges to their respective outer edges, or vice versa, at the dark-mode frequency, and thus the toroidal dipole response is collectively formed.

As a matter of fact, a planar double-ring structure [see Fig. 1(a)] can produce a magnetic resonant mode with antibonding magnetic dipoles (antiparallel magnetic dipoles) in the left and right gaps [17]. This antibonding magnetic-dipole mode, fully different from the routinely adopted magnetic-dipole response in a split-ring resonator, contributes to a Fano-type resonance spectrum, and has been studied in the background of negative refraction as well as EIT [17-20]. In this work, by virtue of the antibonding magnetic-dipole mode, we investigate the possibility of constructing the toroidal dipole moment by rotating the planar double-ring structure into a multifold torus-like metamaterial [Fig. 1(b)].

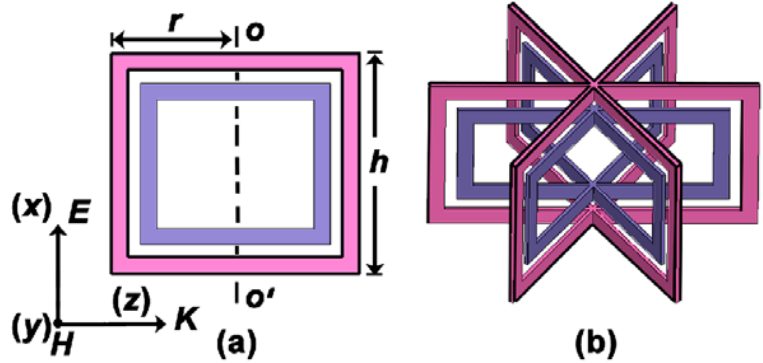

Fig. 1. Unit-cell schematic of the toroidal metamaterial by rotating the double-ring structure with respect to the *oo'* rotation axis. The outer and inner rings are shown in different colors for eye guide. (a) The planar double ring structure (1-fold). (b) The 3-fold double-ring structure with 60-degree intervals between neighboring folds.

## 2. Numerical model for the toroidal metamaterial

Figure 1(a) shows the proposed copper-based structure composed of double rectangular rings with geometric parameters as follows: the gap between the outer and inner rings is 0.2 mm, the metal strip width 0.2 mm, strip thickness 0.02 mm, rotational radius $r = 1.2$ mm, and outer-ring height $h = 4$ mm. The antibonding magnetic-dipole mode in such a double-ring structure has been proposed in our previous work [17], and was subsequently studied by other groups [18-20]. Note that this double-ring unit scale is not in deep subwavelength regarding to the considered wavelength, so that it is not a "homogenizable" metamaterial in a strict sense. To obtain a well-shaped toroidal dipole response visualized by the circularly closed head-to-tail magnetic dipole distribution, the planar double-ring structure is rotated with respect to the symmetry axis to form a multifold torus-like metamaterial [a 3-fold structure shown in Fig.

1(b)]. The intervals for the periodic array of the torus-like metamolecule are 4.4 $mm$ in the *x*-direction and 2.8 $mm$ in the *y*-direction, while only one unit of the metamolecule is considered for the electromagnetic propagation direction. For a realistic consideration, the metallic structure is immerged into a Teflon host medium with relative dielectric constant 2.1 and dielectric loss tangent 0.001. Additionally, in our full-wave simulations based on the finite-element method [21,22], perfect electric and magnetic boundaries are used in compliance with the incident polarization configuration.

### 3. Results and discussions

As for the geometric configuration presented earlier, the antibonding magnetic-dipole mode is clearly shown in Figs. 2(a) and 2(b) around 21 GHz. Figures 2(c) and 2(d) indicate that a perpendicularly arranged double-ring structure does not introduce any irrelevant mode but just bear the same antibonding magnetic-dipole resonance. Figures 2(e) and 2(f) show an obvious formation of the toroidal dipole response in the rotated double-ring metamaterial by increasing the double-ring metamolecule to four folds. As a consequence, a nonzero toroidal dipole moment, oriented along the rotation axis of the multifold torus-like double-ring, is resulted from the closed arrangement of the eight magnetic dipoles with head-to-tail distribution along the torus meridian, though there is vanishing net magnetic dipole in the *yz*-plane [Fig. 2(e)].

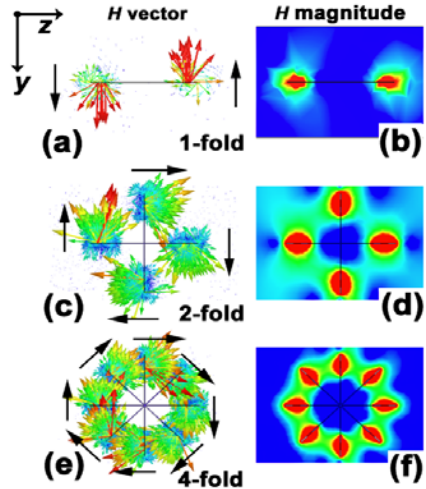

Fig. 2. Magnetic field distributions for torus-like double-ring metamaterials. All plots are in the middle *yz*-plane of the structure. (a) and (b) 1-fold double-ring structure. Antiparallel magnetic dipoles are resonantly confined in the double-ring gap. (c) and (d) 2-fold structure with crossed double rings. Note that the same magnetic resonant mode is shared for the crossed double-ring folds. (e) and (f) 4-fold double-ring structure with a 45-degree rotation interval. The toroidal dipole response with closed head-to-tail magnetic dipole distribution is clearly formed (Black arrows indicate the directions of the magnetic field located around the double-ring gaps).

It should be emphasized that the toroidal dipole response is unlikely induced by the *H*-field component of the polarized incident waves, since the vertically arranged double-ring resonator without any incident *H*-field component perpendicular to its plane also shows the same antibonding magnetic mode [see Fig. 2(c)]. On the contrary, it is considered that the magnetoelectric effect or so-called bianisotropy (i.e., electric coupling to magnetic resonance) is responsible for this toroidal dipole behavior. As a matter of fact, it can be confirmed from Fig. 3 that there are synchronous antiparallel currents induced in each of the inner-and-outer edge pairs of the double-ring metamolecule. This current flowing mode is induced by the electric near-field unbalance between the asymmetric inner and outer edges (i.e., the outer ring is not identical to the inner ring in size) [16,20,23]. This kind of subradiant (or dark) mode excitation due to symmetry-broken geometries usually exhibits a Fano-type spectrum, as was found in literatures [24-28]. Physically, this asymmetric lineshape is caused by the two-pathway interference between the narrow resonant mode and the continuum-like spectrum of the excitation [19,29], rather than an interference result of two resonant modes.

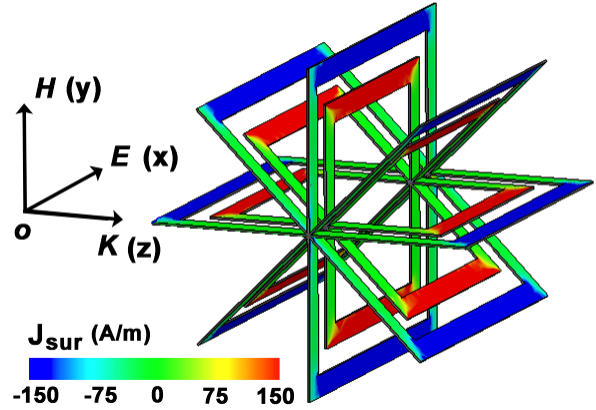

Fig. 3. The magnitude distribution of the induced surface current (*x*-component) at the toroidal dipole resonance. Antiparallel current directions are induced for each of the inner-and-outer edge pairs of the double rings, and thus the toroidal dipole moment is collectively formed.

In the work by T. Kaelberer *et al*., it was confirmed that one of the significant characteristics in the scattering spectrum for a toroidal dipole response is its higher quality factor as compared with that for a magnetic dipole mode [6]. For the Fano-type resonance spectra from the antibonding magnetic-dipole mode in double-ring metamaterials, a resonance dip is always accompanied by a resonance peak in the adjacent frequency, as shown in Fig. 4(a). Moreover, a better shaped torus-like structure with more double-ring folds shows a steeper dip-to-peak profile, which implies a higher quality factor. As far as the transmission peak is concerned, a high quality factor up to 57.6 is obtained for the 4-fold metamolecule [Fig. 4(a)]. This should be attributed to the well-shaped toroidal dipole response, for which case the toroidally confined strong magnetic field is greatly enhanced, as shown in Fig. 2(f). As is well known, a high quality factor can improve the figure of merit in sensing the refractive-index changes of the surrounding medium [27,28]. However, it should be noticed that more folds of double rings will lead to larger transmission suppression due to the increased resonant loss in the multifold metal elements. In addition, the resonance frequency shows a slight shift with the increasing of the rotating folds, from around 20.0 to 23.0 GHz, which originates from the coupling effect between the double-ring folds.

To the last but not the least, although the toroidal dipole response is obtained in this multifold double-ring structure, it should be made clear that if there are any other multipoles, other than the toroidal dipole moment, contributed significantly to the transmission spectra [Fig. 4(a)]. In Fig. 4(b), the scattering powers of various multipoles are calculated by the spatial distribution of current density of an individual molecule with open boundaries [6,7]. Specifically, the *x*-component electric dipoles located at the upper and lower gaps of the double rings (see Fig. 1) will be involved in this resonance, but its scattering cross section [blue line in Fig. 4(b)] is about 20 times weaker than the predominant toroidal moment $T_x$ for the 4-fold structure, at the resonant frequency of 23.0 GHz. For the scattering power by magnetic dipole $M_y$ [black line in Fig. 4(b)], the closed loop distribution makes the scattering field cancel each other, and thus no obvious contributions to the far-field radiation. Moreover, the radiation powers of quadrupoles, i.e., the electric quadrupole $Q_e$ and the magnetic quadrupole $Q_m$, are smaller in strength by 3-4 orders than that scattered by the dipoles.

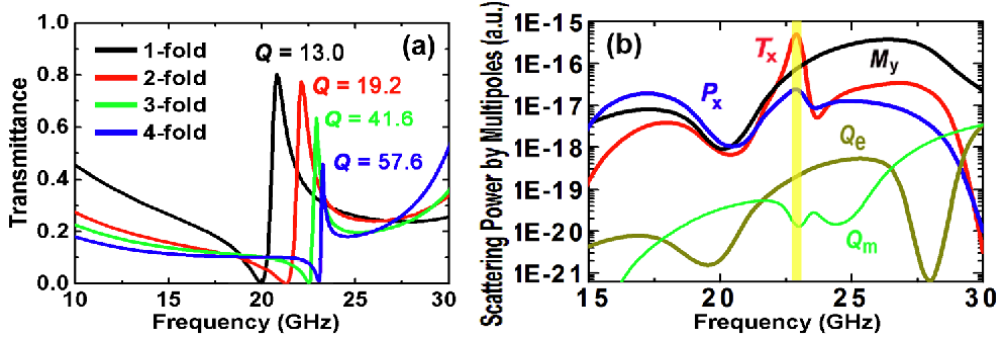

Fig. 4. (a) Transmittance spectra for multifold double-ring metamaterials, where a high quality factor ($Q$) is obtained for the well-shaped torus-like metamolecule (4-fold case). The incident wave propagates along the $z$-direction with polarized electric-field component in the $x$-direction, as specified in Fig. 1. (b) The scattering powers of toroidal dipole ($T_x$), electric dipole ($P_x$), electric quadrupole ($Q_e$), magnetic dipole ($M_y$), and magnetic quadrupole ($Q_m$) calculated from the 4-fold structure. The yellow shadow is for eye guide.

## 4. Summary

In summary, we numerically investigate the toroidal dipole response, represented by the closed distribution of head-to-tail magnetic dipoles, in a torus-like metamaterial composed of multifold double-ring resonators, by virtue of the antibonding magnetic-dipole mode. In essence, the subradiant-mode excitation in symmetry-broken configurations, induced by the nonuniform near-field environment, is responsible for the formation of the intriguing toroidal dipole response. As a consequence, a high quality factor for the resonance spectrum is observed in the well-shaped toroidal metamaterial with 4-fold double rings. Such a metamaterial analogue of the toroidal dipole moment should be helpful in further understandings of the underlying mechanism related to the elusive toroidal electromagnetic response, which will contribute to potential applications regarding to sensing, magnetoelectric effect, and polarization controllability.

**Acknowledgments**

This work was supported by the US National Science Foundation (NSF) Nanoscale Science and Engineering Center CMMI-0751621. DZG also acknowledges the National Natural Science Foundation of China (Nos. 10904012 and 11174051) and the support by Youth Research Plan from SEU.